\documentclass[conference,9pt]{IEEEtran}
\IEEEoverridecommandlockouts

\usepackage{amssymb}
\usepackage{amsmath}
\usepackage{array}
\usepackage{algorithm}
\usepackage{algpseudocode}
\usepackage{booktabs}
\usepackage{color}
\usepackage{dcolumn}
\newcolumntype{d}[1]{D{.}{.}{#1}}
\usepackage[hang,flushmargin]{footmisc}
\usepackage{lipsum}
\usepackage{multirow}
\usepackage[numbers,sort&compress]{natbib}
\usepackage{graphicx}
\usepackage{subfigure}
\usepackage{bm}
\usepackage{threeparttable}

\begin{document}
	
\title{Advances in Microphone Array Processing and Multichannel Speech Enhancement}
	
\name{
	\begin{tabular}{cccc}
Gongping Huang$^1$ & Jesper R.~Jensen$^2$ & Jingdong Chen$^3$  & Jacob Benesty$^4$  \\
Mads G. Christensen$^2$ & Akihiko Sugiyama$^5$ & Gary Elko$^6$ & Tomas Gaensler$^6$  
	\end{tabular}
}
\address{ \large \hskip -7pt \begin{tabular}{c}
$^1$ School of Electronic Information, Wuhan University, 430072, Wuhan, China\\
$^2$ Audio Analysis Lab, Aalborg University, 9220 Aalborg \O st, Denmark. \\
$^3$ CIAIC, Northwestern Polytechnical University, Shaanxi 710072, China\\
$^4$ INRS-EMT, University of Quebec,  Montreal, QC H5A 1K6, Canada \\
$^5$ Damas.cus Corporation, Tokyo, Japan\\ 
$^6$ mh acoustics, USA
\end{tabular}
}

\maketitle
\begin{abstract}
This paper reviews pioneering works in microphone array processing and multichannel speech enhancement, highlighting historical achievements, technological evolution, commercialization aspects, and key challenges. It provides valuable insights into the progression and future direction of these areas. The paper examines foundational developments in microphone array design and optimization, showcasing innovations that improved sound acquisition and enhanced speech intelligibility in noisy and reverberant environments. It then introduces recent advancements and cutting-edge research in the field, particularly the integration of deep learning techniques such as all-neural beamformers. The paper also explores critical applications, discussing their evolution and current state-of-the-art technologies that significantly impact user experience. Finally, the paper outlines future research directions, identifying challenges and potential solutions that could drive further innovation in these fields. By providing a comprehensive overview and forward-looking perspective, this paper aims to inspire ongoing research and contribute to the sustained growth and development of microphone arrays and multichannel speech enhancement.
\end{abstract}


\begin{sloppy}

\section{Introduction}
\label{Sect-Intro}

Microphones are widely integrated into systems such as voice communication and human-machine speech interfaces for sound signal acquisition. Typically, they capture sound waves emitted by sound sources and convert them into electrical signals that can be stored, processed, and transmitted. However, in practice, the signals of interest picked up by microphones are often contaminated by unwanted effects such as additive noise, reverberation, interference, and non-linear distortions, which impair not only the fidelity and quality of the signal of interest but also the speech intelligibility, and performance of automatic speech recognition (ASR).
To mitigate these unwanted effects and enhance speech quality, various speech enhancement methods have been developed. Earliest efforts focused on single-channel scenarios because communication devices in that era were equipped with only a single microphone. While single-channel methods improved the signal-to-noise ratio (SNR) and perceived speech quality, they have a trade-off between noise reduction and speech distortion~\cite{chen2006new}. Furthermore, the signal captured by a single microphone lacks spatial information, making it challenging to distinguish sound sources in the spatial domain and to achieve spatial sound recording and reproduction.

This limitation has led to the development of microphone arrays, where several microphones are arranged into a particular topology to sample the sound field with spatial information. A microphone array typically comprises two main components: a hardware array and a signal processing unit. The hardware array includes a set of microphones and electronic components, such as preamplifiers, A/D converters, memories, and processors. The signal processing unit consists of various algorithms that take the digital microphone signals as input to accomplish specific tasks. Research on the hardware array primarily focuses on designing different array topologies since it is a crucial factor affecting overall performance. In practice, the topology of an array may be constrained by the device in which it is integrated, with common configurations including linear, circular, concentric ring, and spherical arrays~\cite{flanagan1985computer, benesty2008microphone}. To achieve optimal performance, all microphones in the array should exhibit high consistency in terms of sensitivity, dynamic range, frequency response, SNR, and other parameters. Conversely, research in microphone array signal processing primarily focuses on developing algorithms to enhance the signals captured by the array. Based on the correlation between the microphone signals, algorithms can be categorized into signal-independent methods, such as fixed beamforming~\cite{elko2008microphone, benesty2008microphone}, and signal-dependent methods, such as adaptive beamforming~\cite{hoshuyama1999robust}. From the perspective of signal model optimization strategies, they can be classified into model-based methods, which use statistical models and optimization techniques to enhance speech~\cite{gannot2017consolidated, elko2008microphone}, and data-driven methods, which leverage data itself to guide and refine algorithms, such as deep learning-based approaches~\cite{tan2022neural, richter2023speech}.

Microphone arrays, including wireless acoustic microphone networks~\cite{brendel2019distributed}, have the potential to address a variety of acoustic tasks and are utilized in a wide range of applications, such as teleconferencing, hearing aids, and hands-free human-machine interfaces. Beyond these traditional applications, microphone arrays are increasingly being applied in diverse fields, including biomonitoring, surveillance for security, and robot auditions~\cite{richard2023audio}.
Over the past few decades, the rapid and significant development of microphone arrays and multichannel speech enhancement methods has led to their increasingly important role and the emergence of a growing number of applications~\cite{elko2008microphone, benesty2023microphone}. In particular, the last ten years have witnessed significant advancements in deep learning, which have driven a shift in microphone arrays and multichannel speech enhancement towards data-driven methods based on deep learning~\cite{zmolikova2023neural, heymann2016neural, xiao2016deep}.

This paper aims to provide a brief overview of microphone arrays and multichannel speech enhancement methods, reviewing pioneering works, and highlighting key aspects such as historical achievements, technological evolution, commercialization, and significant challenges and datasets. It offers insights into the progression of these fields, aiming to inspire ongoing research and contribute to the continued growth and development of microphone arrays and multichannel speech enhancement.


\section{Foundational Developments}
\label{Sect-FD-MA}

A microphone array captures sound signals using multiple microphones arranged in a specific spatial topology and applies signal processing techniques to enhance signals from particular sound sources. Early research in this area was heavily influenced by sensor array theory from radar and sonar signal processing. However, it soon became clear that the challenges faced by microphone arrays are markedly different from those encountered in radar and sonar due to the following several distinct and significant reasons~\cite{benesty2015design}.
\begin{list}{$\bullet$ \hfill}
{\leftmargin=3ex \labelwidth=1.5ex  \labelsep=0.5ex \topsep=0.5ex \itemsep=0.5ex \parsep=0ex \usecounter{enumi}}
\item Microphone arrays handle speech and audio signals, which are broadband in nature with a frequency range of approximately 20 Hz to 20 kHz. Developing algorithms that perform consistently across such a wide frequency range is extremely challenging.
    
\item Microphone arrays often need to estimate the statistical properties of signals. However, the highly non-stationary nature of speech and audio signals makes this estimation particularly challenging.

\item Microphone arrays often operate in reverberant environments with a large number of reflections, which can significantly degrade the performance of processing algorithms.
    
\item Due to the physical and computational restrictions in commercial products, the number of microphone is usually limited, making it challenging to achieve good spacial directivity.
\end{list}
Due to these reasons, many algorithms that perform well in radar and sonar sensor array applications often fail to deliver satisfactory results with microphone arrays. Consequently, enhancing the effectiveness of microphone arrays and their associated processing algorithms has been a focal point of research over the past few decades.

Typically, microphone arrays consist of multiple microphones arranged in fixed patterns, such as linear or circular configurations. These arrangements utilize the spatial diversity of the microphones to design filters that enhance the desired source signal~\cite{brandstein2001microphone, huang2022fundamental}. Due to hardware constraints, early microphone array beamforming often employed delay-and-sum (DS) beamforming, a straightforward method for achieving directionality~\cite{flanagan1985computer}. However, the beamwidth of a DS beamformer is inversely proportional to the frequency, making the mainlobe wider at lower frequencies and narrower at higher frequencies. This characteristic reduces its effectiveness in handling noise and interference, especially at low frequencies. Furthermore, the frequency-dependent response of the DS beamformer can unintentionally act as a low-pass filter if the signal incidence angle deviates from the steering direction, and it does not consistently suppress noise across the entire frequency spectrum. As a result, designing a broadband beamformer that delivers consistent performance over a wide frequency range is both complex and challenging.

One proposed solution to address this challenge is to divide the array into subarrays, each designed to cover a different frequency band. 
This creates a microphone array composed of multiple harmonically-nested subarrays, allowing it to handle a broad bandwidth of interest~\cite{flanagan1985computer, flanagan1991autodirective}. However, these nested subarrays can become quite large and may not always be practical for real-world applications.
An alternative solution is phase-based time-frequency filtering~\cite{aarabi2004phase} with a constant beamwidth~\cite{sugiyama2015directional}, which is particularly useful for personal computer applications. Another effective method for achieving consistent broadband performance is the differential beamformer. This approach is widely used in practical systems due to its ability to deliver high directional gains and create frequency-invariant beampatterns with compact microphone arrays~\cite{elko1996microphone, elko2000superdirectional, elko2004differential}. The concept of differential microphone arrays (DMAs) is inspired by the directional ribbon microphone, which measures acoustic particle velocity or pressure difference~\cite{olson1946gradient, weinberger1933uni}. A notable limitation of early directional microphones was their lack of adaptability in adjusting their directional patterns to meet diverse requirements; their directional response was essentially fixed at the time of manufacture. Some microphones, however, offered multiple selectable patterns, such as the Western Electric 639, which featured both ribbon and dynamic elements, allowing users to choose from six different patterns.
The modern concept of DMAs was developed using digital signal processing techniques, where a number of pressure microphones are arranged into a particular geometry so the differentials of the acoustic pressure field can be measured by combining the outputs of a number of omnidirectional microphones. A new method to design differential beamformers in the short-time-Fourier-transform (STFT) domain based on null-constraints is further proposed, which offers significant advantages in design flexibility and robustness, i.e., improving the white noise amplification problem~\cite{benesty2012study, benesty2016fundamentals}. A generalization of differential beamforming to circular microphone arrays~\cite{benesty2015design, huang2017design}, concentric circular microphone arrays~\cite{huang2018insights}, and rectangular arrays has been proposed~\cite{benesty2023microphone}, which offers further advantages like enhanced beam steering capabilities and increased robustness~\cite{ jin2022differential}.

A particularly representative microphone array beamforming method is the spherical harmonics-decomposition based approach, which involves decomposing the sound field into its spherical harmonic components~\cite{meyer2002highly}. Instead of forming beampatterns through the traditional integration of filtered signals from individual microphones, this method utilizes spatially decomposed eigenbeams to achieve beamformer design. The Eigenmike® spherical microphone array represents a significant milestone in this microphone array technology~\cite{Eigenmike2013}. The Eigenmike consists of two principal components: the eigenbeamformer, which transforms microphone signals into an orthonormal spherical harmonics, and the modal beamformer, which combines these spherical harmonics to generate the desired beampattern. The advanced design and robust capabilities of the Eigenmike have made it a cornerstone in the development of microphone array technology and have garnered widespread adoption among researchers.

\section{Recent Advancements and Cutting-edge Research}
\label{Sect-RACR-MA}

With the rapid evolution of deep learning, much like other areas within the research field of audio and acoustic signal processing (AASP)~\cite{richard2023audio}, the domain of microphone arrays and multichannel speech enhancement is experiencing a shift towards data-driven approaches based on deep learning~\cite{wang2018supervised}. Over the past decade, this trend has propelled significant advancements, resulting in the development of numerous deep neural network (DNN)-based techniques. The current literature on these deep learning based data-driven approaches can primarily be categorized into the following three main types.

\begin{list}{$\bullet$ \hfill}
{\leftmargin=3ex \labelwidth=1.5ex  \labelsep=0.5ex \topsep=0.5ex \itemsep=1ex \parsep=0ex \usecounter{enumi}}
\item 
{\it Utilizing DNNs for Statistical Estimation in Beamforming}:
DNNs have been essential in statistical estimation methods, which were among the pioneering approaches in this domain~\cite{tao2024learning}. These methods diverge from conventional techniques that rely on voice activity detection (VAD) or other strategies to estimate the statistical properties of signals~\cite{tao2024learning}. Initially, deep learning applications in this domain involved mask-based methods, aiming to improve speech enhancement by applying a mask to the spectrogram of the noisy input signal to isolate the desired speech components and calculating relevant statistics. Techniques like the ideal binary mask and ideal ratio mask provide binary or ratio masks indicating whether a given time-frequency bin contains speech or noise, with neural networks trained to approximate these ideal masks~\cite{heymann2016neural}.
By employing DNNs, they estimate essential statistics for traditional beamforming, such as the covariance matrix of noise signals or the relative transfer functions (RTFs) of the desired signal. The DNNs enhance the accuracy of these estimates by learning the characteristics of noise and signals within the data, thereby improving the performance of the beamformer~\cite{higuchi2016robust}.

\item 
{\it Direct Estimation of Beamforming Weights via DNNs}:
In this category, DNNs are trained to directly learn and estimate the coefficients of multichannel filter weights employed in beamforming. By optimizing these weights through training, the DNNs achieve speech signal enhancement without the need to estimate the statistical properties of the noise or clean signals~\cite{xiao2016deep}. Among the advanced techniques, all-neural beamformers integrate deep learning directly into the beamforming process, optimizing spatial filtering and noise reduction in a single, end-to-end framework~\cite{tan2022neural}.

\item 
{\it Direct Processing of Multichannel Data Using DNNs}:
As the field progressed, researchers shifted towards end-to-end learning frameworks that operate directly on the raw audio waveform, learning to map noisy input signals to clean output signals. 
These approaches apply deep learning networks directly to process multichannel data, bypassing the traditional structure of beamformers. The DNNs enhance speech by processing the raw multichannel signals, often involving end-to-end learning frameworks to process audio signals without the need for explicit feature extraction, effectively capturing temporal dependencies~\cite{pandey2022tparn}.
\end{list}

There are also hybrid methods that combine features of the aforementioned techniques. For instance, some methods may utilize DNNs in the process of estimating statistical properties while also optimizing filter weights, or they might integrate traditional beamforming techniques within an end-to-end learning framework. These advancements have transformed traditional approaches, introducing more sophisticated, accurate, and efficient methods for handling complex audio environments. 
Additionally, attention-based models, such as {\it transformer} networks, have been applied to microphone array processing to better capture dependencies across different channels and improve the focus on relevant sound sources~\cite{pandey2022tparn}. Recent advancements have also introduced diffusion models and generative approaches, leveraging the power of generative frameworks for enhanced speech processing~\cite{richter2023speech}. 
Current trends in the field also include multimodal approaches that combine audio data with visual or other sensory inputs to improve speech enhancement and target signal extraction~\cite{michelsanti2021overview}.

For multichannel speech enhancement, a critical issue lies in the construction of training data. In existing methods, multichannel signals primarily rely on simulation generation, where multichannel room impulse responses (RIRs) are created using the image model method. The single-channel clean source signals are from various open datasets and noise is typically derived from single-channel noise datasets or generated directly by computers as white or spatially diffuse noise. These single-channel clean signals are then convolved with RIRs to produce multichannel signals, which are finally combined with noise to obtain the observation signals.
However, this simulation-based approach to generating multichannel signals faces the following several challenges.

\begin{list}{$\bullet$ \hfill}
{\leftmargin=3ex \labelwidth=1.5ex  \labelsep=0.5ex \topsep=0.5ex \itemsep=1ex \parsep=0ex \usecounter{enumi}}
\item 
The image model for generating RIRs is confined to the representation of standard rectangular rooms~\cite{allen1979image}. Real-world environments, however, are considerably more complex, with room shapes potentially non-standard and various obstacles within the room with varying reflection coefficients. Consequently, image models may not precisely replicate the conditions encountered in actual environments.
\item 
Modeling multichannel signals by convolving clean source signal with RIRs does not perfectly align with reality. The convolution process usually assumes that the acoustic channel remains unchanged over a period, whereas in real environments, sources may be directional and moving, and the acoustic channel is often time-varying.
\item 
Modeling multichannel signals with additive noise may not accurately reflect real-world situations, such as those encountered in cocktail party scenarios. This could be one reason why multichannel speech enhancement methods still fall short of expectations and have not achieved the same level of success as their single-channel counterparts in real-world applications.
\end{list}

For deep learning-based multichannel speech enhancement, the selection of the target signal is not yet fully defined. In practice, the choice of target signal is typically categorized into two categories: 1) the clean source (direct-path) signal; and 2) the direct-path signal plus early reflections observed at the reference microphone. Theoretically, if the clean source signal could be completely recovered, optimal speech enhancement could be achieved. However, in real reverberant environments, the energy of the direct-path signal constitutes a relatively small proportion of the observed signal, making the complete recovery of the direct-path signal significantly challenging. Moreover, not all reflections are detrimental to speech perception. It is commonly believed that early reflections may be beneficial to speech quality and intelligibility (the first $20$~ms are mostly integrated by the ear and typically increase the beneficial signal power).
In contrast, late reflections, due to their incoherence and extended duration beyond perceptual integration times, can degrade speech quality and intelligibility. Thus, the objective often becomes to estimate the direct-path signal along with early reflections. Existing methods typically define early reflections as those occurring within less than 50 ms from the direct sound. This threshold is based on early psychoacoustic studies and empirical evidence. However, the impact of different parts of the reverberant signal on speech perception can vary across different tasks and scenarios. Therefore, determining the appropriate target signal and developing the most effective data-driven learning model for multichannel speech enhancement remains an important area of investigation.

While deep learning techniques have markedly advanced speech and audio processing over the past decade, they often overlook the intrinsic physical aspects of the task. Furthermore, purely deep learning-based methods tend to be unexplainable and lack interpretability. A promising approach to overcoming these limitations is to incorporate prior knowledge of the data and underlying physics into the models~\cite{raissi2019physics}. This involves combining deep learning with traditional signal processing to develop physically-informed models that enhance generalization across different conditions, environments, and array configurations.
The deep learning models should also incorporate perceptual metrics into their loss functions, as human perception is influenced by more than just physics and algorithmic architectures.

\section{Enabling Factors and Key Contributions}
\label{Sect-EFKC-MA}


Publicly available datasets and organized challenges are crucial for advancing research in microphone arrays and multichannel speech enhancement. These resources provide standardized benchmarks, enabling researchers to develop, test, and compare algorithms under consistent conditions.
One dataset in this field is the Computational Hearing in Multisource Environments (CHiME) series, which has been widely used for evaluating speech enhancement and recognition algorithms in noisy environments~\cite{yoshioka2015ntt}. 
Since its inception in 2011, the CHiME series has held seven editions, establishing itself as an important platform for advancing robust speech processing technologies. 
The Augmented Multi-party Interaction (AMI) corpus, containing recordings of meetings, captured with multiple microphone arrays~\cite{kraaij2005ami}. The REverberant Voice Enhancement and Recognition Benchmark (REVERB) Challenge focuses on speech enhancement and recognition in reverberant conditions~\cite{kinoshita2013reverb}. The MIRaGe dataset features multichannel recordings made in an acoustic lab with adjustable reverberation time~\cite{vcmejla2021mirage}.
The Distant-speech Interaction for Robust Home Applications (DIRHA) dataset offers recordings in real home environments with multiple microphone arrays. The SPeech Enhancement for Augmented Reality (SPEAR) Challenge focuses on improving speech enhancement techniques for augmented reality applications. The Audio-Visual Speech Enhancement (AVSE) Challenge combines audio and visual data to enhance speech signals. The Clarity Enhancement Challenge (CEC3) focuses on improving the performance of hearing aids for speech-in-noise scenarios. The L3DAS22 Challenge emphasize the spatial aspects of sound and promoting advancements in three-dimensional audio technologies. 

These datasets and challenges not only facilitate benchmarking of new algorithms but also promote reproducibility and transparency in research. However, despite the significant contributions of these datasets to the field, there remains a notable absence of a standard, authoritative, and widely recognized dataset akin to ImageNet in the image domain. As new datasets and challenges continue to emerge, they will provide fresh opportunities for innovation in microphone array technology and its applications.
Moreover, many research studies publish improvements in certain metrics that could be incidental, with the methodologies often inadequately described. This underscores the importance of publicly sharing data and code to enhance reproducibility and ensure that findings can be verified and built upon.


Performance metrics are essential for evaluating and comparing the performance of microphone arrays processing methods. Early metrics were often simplistic, primarily measuring SNR improvements or noise reduction. Some more comprehensive metrics are developed to better reflect real-world performance, such as the Short-Time Objective Intelligibility (STOI)~\cite{Taal2010shorttime}, which assesses speech intelligibility despite noise or distortions, Perceptual Evaluation of Speech Quality (PESQ) and Perceptual Objective Listening Quality Assessment (POLQA), which evaluate the overall quality of processed speech, considering factors such as clarity and naturalness. 

These metrics have contributed to research consistency by providing common benchmarks for evaluating and comparing different algorithms. Despite these advancements, several challenges remain. While these metrics are valuable for establishing benchmarks, those designed for single-channel scenarios may not fully capture the complexities of spatial audio processing and multichannel environments. Additionally, these metrics can have limitations; for example, improvements in STOI do not always correlate with better results in listening experiments~\cite{gelderblom2023predictive}. Some studies suggest that it may be worthwhile to explore non-intrusive quality and intelligibility measures, though these are not yet highly successful, as well as metrics suitable for in-the-loop optimization and validation, given that most current metrics are primarily used for validation~\cite{deoliveira2024PESQetarian}. The continuous refinement and development of metrics will be essential for accurately assessing and advancing microphone array technologies.

\section{Critical Applications}
\label{Sect-CA-MA}

Microphone arrays have become essential in many applications. In teleconferencing systems, microphone arrays play a crucial role in ensuring clear and intelligible communication. Compared to using a gooseneck microphone close to the talker, using microphone array beamforming can offer significant directional gain and more effective noise reduction, improving the overall audio experience for participants. Microphone arrays can also track multiple speakers and switch focus dynamically, making them ideal for conference rooms, where participants may move or speak from different locations. 
Hearing aids can also benefit from advancements in microphone array technology~\cite{kates1996comparison}. Given the stringent size and power consumption constraints in hearing aids, compact microphone arrays offer significant potential. Modern hearing aids that incorporate microphone arrays and leverage deep learning are more effective at distinguishing between speech and background noise, thereby improving the user's ability to comprehend conversations in noisy environments.
Smart home devices, such as voice-activated assistants and smart speakers, rely heavily on microphone arrays for far-field voice capture, enabling them to recognize and respond to voice commands. Notable examples like Amazon Echo, Apple HomePod, and Xiaomi Mi AI Speaker utilize advanced microphone arrays and multichannel processing algorithms to ensure high-quality speech acquisition.

Emerging applications include autonomous vehicles and virtual reality (VR)/augmented reality (AR) applications. Microphone arrays are primarily utilized to enhance in-car voice assistants and interaction systems, facilitating seamless communication between the driver and the system in the vehicle. These arrays enable accurate voice command recognition, allowing for efficient control of navigation, entertainment, and other in-car functions. In VR and AR applications, microphone arrays capture high-fidelity spatial audio by recording sound from multiple directions, which is essential for creating an immersive user experience.  This promising application is anticipated to attract considerable interest in future consumer devices.

\section{Conclusions and Perspectives}
\label{Sect-Con}

Microphone arrays have made remarkable progress and are now widely adopted in various fields. Over the past decade, the rapid evolution of deep learning has shifted microphone arrays and multichannel speech enhancement toward data-driven approaches, attracting increasing research attention. Despite this progress, deep learning-based microphone array speech enhancement methods face numerous challenges, such as generating sufficient multichannel training data to cover diverse environments. The development of a standard dataset is eagerly anticipated to enable easy and fair comparisons of algorithms. Designing systems for diverse users with unique auditory requirements and environments presents another challenge. The computational load and large-scale data required for deep learning must also be addressed, balancing high performance with minimal computational load, especially for real-time applications like mobile communication and hearing aids. Enabling microphone arrays to achieve true spatial hearing, focusing on one or several specific sounds much like the human auditory system, remains an ongoing effort. 

Before concluding this paper, it is important to note that, due to space constraints, this review is neither exhaustive nor comprehensive. Many significant papers, books, and other publications in the field may not be mentioned or cited. Readers interested in the topic are encouraged to explore the extensive range of papers, edited volumes, and monographs available in the field.
	
\footnotesize
	
\bibliographystyle{ieeetr}

\end{sloppy}

\end{document}